\documentstyle[amssymb,preprint,aps]{revtex}

\begin{document}
\title{T-dual R-R zero-norm states, D-branes and S-duality of type II string theory}
\author{Jen-Chi Lee\thanks{
e-mail: jcclee@cc.nctu.edu.tw}}
\address{Department of Electrophysics, National Chiao Tung University, Hsinchu 30050,%
\\
Taiwan, R. O. C.}
\maketitle

\begin{abstract}
We calculate the R-R zero-norm states of type II string spectrum. To fit
these states into the right symmetry charge parameters of the gauge
transformations of the R-R tensor forms, one is forced to T-dualize some
type I open string space-time coordinates and thus to introduce D-branes
into the theory. We also demonstrate that the constant T-dual R-R 0-form
zero-norm state, together with the NS-NS singlet zero-norm state are
responsible for the SL(2,Z) S-duality symmetry of the type II B string
theory.


\end{abstract}


\section{Introduction}

\label{sec:intro}It has been pointed out for a long time that the complete
space-time symmetry\cite{1} of string theory is related to the zero-norm
state (a physical state that is orthogonal to all physical states including
itself) in the old covariant quantization of the string spectrum.\cite{2}
This observation had made it possible to explicitly construct many stringy ($%
\alpha ^{\prime }\rightarrow \infty $) massive symmetry of the theory. This
includes the $w_{\infty }$ symmetry of the toy 2D string\cite{3} and the 
{\it discrete} massless and massive T-duality symmetry of closed bosonic
string.\cite{4} The authors of \cite{5} show that, in string theory, some
target space mirror symmetry of N=2 backgrounds on group manifolds is a
Kac-Moody gaugy symmetry. Thus, like T-duality, it should be related to the
zero-norm states. On the other hand, the massless and massive SUSY, and some
new enlarged spacetime boson-fermion symmetries induced by zero-norm states
were also discussed in \cite{6}. It is thus of interest to study the R-R
zero-norm state and its relation to D-brane which was recently shown by
Polchinski to be the symmetry charge carrier of the propagating R-R forms.%
\cite{7}

Presumably, there should be no R-R zero-norm state in the type II string
spectrum since the fundamental string does not interact with the R-R forms.
However, to our surprise, it was discovered that there do exist both
massless and massive R-R zero-norm states in the type II string spectrum.%
\cite{6} It was then realized that the degree of freedom of massless R-R
zero-norm states does not fit into that of the symmetry parameters of the
propagating R-R forms and thus resolved the seeming inconsistency. This
observation gives us another justification of the well-known wisdom that
perturbative string does not carry the massless R-R charges, although the
existence of these R-R zero-norm states remain mysterious.

In this paper, we will show that the T-dual R-R zero-norm states serve as
the right symmetry parameters of the gauge transformations of the R-R
propagating forms. Also one is forced to introduce Type I open string and
D-branes into the type II string theory to incorporate these T-dual R-R
zero-norm states. Our {\it spacetime} zero-norm states argument here is in
complementary with the {\it worldsheet} string vertex operator argument
first given by Binnchi, Pradisi and Sagnotti.\cite{8} They considered R-R
one-point function in the (-$\frac{1}{2}$,-$\frac{3}{2}$) ghost picture on
the {\it disk} and resulted in a conclusion which was consistent with
D-brane as R-R charge carrier\cite{7}. As an important application, we
demonstrate that the constant T-dual R-R 0-form zero-norm state, together
with the NS-NS singlet zero-norm state which was always neglected in the
previous discussions, are responsible for the {\it discrete} SL(2,Z)
S-duality symmetry of the type II B string theory.\cite{9} This discovery
suggests that not only stringy ($\alpha ^{\prime }\rightarrow \infty $)
symmetry but also strong-weak ($g_{s}\rightarrow \infty $) duality symmetry
are related to the existence of zero-norm states of the spectrum.

\section{T-dual R-R zero-norm states}

The massless physical NS state of the open superstring is (we use the
notation in Ref \cite{10})

\begin{equation}
\varepsilon _{\mu }b_{-\frac{1}{2}}^{\mu }\left| 0,k\right\rangle ;\text{ }%
k\cdot \varepsilon =0,\text{ }k^{2}=0
\end{equation}
In addition, there is a singlet zero-norm state

\begin{equation}
k_{\mu }b_{-\frac{1}{2}}^{\mu }\left| 0,k\right\rangle ;\text{ }k^{2}=0\text{
.}
\end{equation}
The NS-NS symmetries of graviton and antisymmetry tensor of type II string
were derived through the following two zero-norm states

\begin{eqnarray}
&&\varepsilon _{\mu }b_{-\frac{1}{2}}^{\mu }\left| 0,k\right\rangle \otimes
k_{\mu }\stackrel{\sim }{b}_{-\frac{1}{2}}^{\mu }\left| 0,k\right\rangle , 
\nonumber \\
&&k_{\mu }b_{-\frac{1}{2}}^{\mu }\left| 0,k\right\rangle \otimes \varepsilon
_{\mu }\stackrel{\sim }{b}_{-\frac{1}{2}}^{\mu }\left| 0,k\right\rangle
\end{eqnarray}
in the first order weak field approximation (WFA)\cite{11}. The remaining
interesting singlet zero-norm state

\begin{equation}
k_{\mu }b_{-\frac{1}{2}}^{\mu }\left| 0,k\right\rangle \otimes k_{\mu }%
\stackrel{\sim }{b}_{-\frac{1}{2}}^{\mu }\left| 0,k\right\rangle
\end{equation}
will be discussed in the next section.

We now discuss the massless R state. The only propagating spinor is

\begin{equation}
\left| \stackrel{\rightharpoonup }{S},k\right\rangle u_{\stackrel{%
\rightharpoonup }{s}};\text{ }F_{0}\left| \stackrel{\rightharpoonup }{S}%
,k\right\rangle u_{\stackrel{\rightharpoonup }{s}}=0\text{ .}
\end{equation}
The GSO operator in the massless limit reduces to the chirality operator,
and only one of the chiral spinor $8_{s}($or $8_{c})$ will be projected out.
In addition, there is a massless fermionic zero-norm state

\begin{equation}
k_{\mu }\Gamma _{\stackrel{\rightharpoonup }{s},\stackrel{\rightharpoonup }{s%
}}^{\mu }\left| \stackrel{\rightharpoonup }{S},k\right\rangle \theta _{%
\stackrel{\rightharpoonup }{s}}\text{ .}
\end{equation}
Eq.(6) is the only massless solution of the following

\begin{equation}
F_{0}\left| \psi \right\rangle ,\text{where }F_{1}\left| \psi \right\rangle
=L_{0}\left| \psi \right\rangle =0\text{.}
\end{equation}
The state in equation (6) is crucial in the discussion of this paper. Note
that $k\cdot \Gamma \left| \stackrel{\rightharpoonup }{S},k\right\rangle
\theta _{\stackrel{\rightharpoonup }{s}}$ is left-handed if $\left| 
\stackrel{\rightharpoonup }{S},k\right\rangle \theta _{\stackrel{%
\rightharpoonup }{s}}$ is right-handed and both spinors have exactly the
same degree of freedom. The massless propagating R-R states of type II
string consist of tensor forms

\begin{equation}
G_{\alpha \beta }=\sum_{k=0}^{10}\frac{i^{k}}{k!}G_{\mu _{1}\mu _{2}...\mu
_{k}}(\Gamma ^{\mu _{1}\mu _{2}...\mu _{k}})_{\alpha \beta }\text{,}
\end{equation}
where $\Gamma ^{\mu _{1}\mu _{2}...\mu _{k}}$ are the antisymmetric products
of gamma-matrix, and $\alpha ,\beta $ are spinor indices. There is a duality
relation which reduces the number of independent tensor components to up to
k=5 form. The on-shell conditions, or two massless Dirac equations, imply G
is indeed a field strength and can be written as

\begin{equation}
G_{(k)}=dA_{(k-1)}
\end{equation}
which means perturbative string states do not carry the {\it massless} R-R
symmetry charges. We are now in a position to discuss the R-related symmetry
charges. Let's first introduce the NS-R (R-NS) SUSY zero-norm states\cite{6}

\begin{equation}
\text{ \qquad }k\cdot b_{-\frac{1}{2}}\left| 0,k\right\rangle \otimes \left| 
\stackrel{\rightharpoonup }{S},k\right\rangle \overline{u}_{\stackrel{%
\rightharpoonup }{s}}\text{and }\left| \stackrel{\rightharpoonup }{S}%
,k\right\rangle u_{\stackrel{\rightharpoonup }{s}}\otimes k\cdot \widetilde{b%
}_{-\frac{1}{2}}\left| 0,k\right\rangle
\end{equation}
for the II A theory and a trivial modification for the II B theory. The
corresponding worldsheet vertex operator in the ($0,-\frac{1}{2}$) picture
for say the first state in equation (10) is

\begin{eqnarray}
&&k_{\mu }(\partial x^{\mu }(z)+ik\cdot \psi \psi ^{\mu })e^{ik\cdot
x(z)}u_{\alpha }\stackrel{\sim }{S}^{\alpha }(\overline{z})e^{-\frac{1}{2}%
\stackrel{\sim }{\phi }}e^{ik\cdot x(\overline{z})}  \nonumber \\
&=&\partial e^{ik\cdot x(z)}u_{\alpha }\stackrel{\sim }{S}^{\alpha }e^{-%
\frac{1}{2}\stackrel{\sim }{\phi }}e^{ik\cdot x(\overline{z})},
\end{eqnarray}
which is a worldsheet total derivative and, as in the case of bosonic sector%
\cite{2}, one can introduce a worldsheet generator and deduce the SUSY
current to be

\begin{equation}
\stackrel{\sim }{Q}_{\alpha ,-\frac{1}{2}}=\stackrel{\sim }{S}^{\alpha }e^{-%
\frac{1}{2}\stackrel{\sim }{\phi }},
\end{equation}
where $\stackrel{\sim }{S}^{\alpha }$and $\stackrel{\sim }{\phi }$ are the
right-moving spin field and the bosonized superconformal ghost respectively.
This {\it zero-norm state derivation }is consistent with the original
approach.\cite{12} The advantage of our approach is that one can generalize
to derive the enlarged stringy boson-fermion symmetry by using the massive
fermion zero-norm state of the spectrum. We give one example here. There
exists a m=2 NS-R zero-norm state

\begin{equation}
\left[ 2\theta _{\mu \nu }\alpha _{-1}^{\mu }b_{-\frac{1}{2}}^{\nu
}+k_{[\lambda }\theta _{\mu \nu ]}b_{-\frac{1}{2}}^{\lambda }b_{-\frac{1}{2}%
}^{\mu }b_{-\frac{1}{2}}^{\nu }\right] \left| 0,k\right\rangle \otimes 
\widetilde{\alpha }_{-1}^{\lambda }\left| \stackrel{\rightharpoonup }{S}%
,k\right\rangle u_{\lambda ,\stackrel{\rightharpoonup }{s}}
\end{equation}
with $\theta _{\mu \nu }=-\theta _{\nu \mu },$ $k^{\mu }\theta _{\mu \nu }=0$
and

\begin{eqnarray}
\lbrack (k\cdot d_{0})\alpha _{-1}^{\mu }+d_{-1}^{\mu }]u_{\mu ,\stackrel{%
\rightharpoonup }{s}} &=&0,  \nonumber \\
d_{0}^{\mu }u_{\mu ,\stackrel{\rightharpoonup }{s}} &=&0.
\end{eqnarray}
The corresponding vertex operator is calculated to be

\begin{eqnarray}
&&[2\theta _{[\mu \nu ],\lambda \alpha }(\partial x^{\mu }\partial x^{\nu
}-\psi ^{\mu }\partial \psi ^{\nu }+ik\cdot \psi \psi ^{\mu }\partial x^{\nu
})+k_{[\delta }\theta _{\mu \nu ],\lambda \alpha }(3\partial x^{\mu
}+ik\cdot \psi \psi ^{\mu })  \nonumber \\
&&\psi ^{\nu }\psi ^{\delta }]\overline{\partial }x^{\lambda }k\cdot 
\overline{\psi }e^{-\frac{1}{2}\stackrel{\sim }{\phi }}\stackrel{\sim }{S}%
^{\alpha }e^{ik\cdot x(z,\overline{z})}
\end{eqnarray}
where $\theta _{\mu \nu ,\lambda \alpha }\equiv \theta _{\mu \nu }\cdot
u_{\lambda \alpha }.$ It is straight-forward to construct the corresponding
ward identity although the symmetry transformation law of the background
fields is not easy to write down at this point.

We now turn to discuss the R-R zero-norm states. For the massless level, we
have the following zero-norm states

\begin{equation}
k_{\mu }\Gamma _{\stackrel{\rightharpoonup }{s}^{\prime }\stackrel{%
\rightharpoonup }{s}}^{\mu }\left| \stackrel{\rightharpoonup }{S}%
,k\right\rangle \theta _{\stackrel{\rightharpoonup }{s}}\otimes \left| 
\stackrel{\rightharpoonup }{S},k\right\rangle u_{\stackrel{\rightharpoonup }{%
s}}\text{ \qquad (II A)}
\end{equation}
and

\begin{equation}
k_{\mu }\Gamma _{\stackrel{\rightharpoonup }{s}^{\prime }\stackrel{%
\rightharpoonup }{s}}^{\mu }\left| \stackrel{\rightharpoonup }{S}%
,k\right\rangle \overline{\theta _{\stackrel{\rightharpoonup }{s}}}\otimes
\left| \stackrel{\rightharpoonup }{S},k\right\rangle u_{\stackrel{%
\rightharpoonup }{s}}\text{ \qquad (II B).}
\end{equation}

These are tensor forms as in equation (8). The on-shell condition on the
right mover together with the trivial identity ($k\cdot \Gamma $)$^{2}\left| 
\stackrel{\rightharpoonup }{S},k\right\rangle \theta _{\stackrel{%
\rightharpoonup }{s}}=0$ on the left mover imply, as in equation (9), that

\begin{equation}
F_{(k)}=d\omega _{(k-1)}.
\end{equation}
Note that, for the II A (II B) theory, $\omega _{(p)}$ in eq(18) does not
fit into the gauge symmetry parameters of $A_{(p)}$ forms of II A (II B)
theory in eq(9) since they share the same tensor index structures. In fact,
for a $p+1$ form $A_{(p+1)}$, one needs a $p$ form $\widetilde{\omega _{(p)}}
$ symmetry parameters, as can be seen from its spacetime coupling to D-brane

\begin{equation}
\int_{\text{world vol of D-brane}}A_{(p+1)}\equiv \int d^{p+1}\xi A_{\mu
_{1}\mu _{2}...\mu _{p+1}}(x)\partial _{1}x^{\mu _{1}}\cdot \cdot \cdot
\partial _{p+1}x^{\mu _{p+1}},
\end{equation}
which implies a space-time gauge symmetry

\begin{equation}
A_{(p+1)}\rightarrow A_{(p+1)}+d\widetilde{\omega }_{(p)}.
\end{equation}
This justifies that no perturbative type II string state carries the R-R
charge. On the other hand, it is well-known that each time we T-dualize in
an additional direction the dimension of the D-branes goes down by one and
the R-R forms lose an index. To include the right closed string zero-norm
state $\widetilde{\omega }_{(p)}$, one is thus forced to introduce the type
I {\it unoriented} open string and T-dualizes k = odd (even) numbers of
space-time coordinates and then takes the noncompact limit $R\rightarrow 0$
for each compatified radius. For k = odd (even), one has type II A (II B)
string states {\it in the bulk far away from D-branes}. The right $%
\widetilde{\omega }_{(p)}\equiv \omega _{(p+1)}^{(T)}$ state, the T-dual R-R
zero-norm state, is thus attached to the D p-brane for p= even (odd) in II A
(II B) theory. Note that, near the D-branes, the orientation projection of
the Type I theory leaves only one linear combination of two SUSY
charges(SUSY zero-norm states in eq.(10)) of the Type II theory in the bulk.
It is $Q_{\alpha }^{^{\prime }}+(\Pi _{m}^{k}\beta ^{m}\stackrel{\sim }{%
Q^{\prime }})_{\alpha }$ with $\beta ^{m}\equiv \Gamma ^{m}\Gamma .$ The
T-dual R-R zero-norm states attached in the boundary of the open string
1-loop diagram with D-branes are the T-dual version of the R-R zero-norm
states in the bulk of the closed string tree diagram. Our argument resolves
the puzzle of seeming unwanted R-R zero-norm states in perturbative type II
string spectrum and simultaneously {\it motivates the introduction of
D-branes into the theory }which is complementary to the argument in Ref\cite
{7}. The space-time T-dual R-R zero-norm state has an interesting analogy
from worldsheet vertex operator point of view. The authors of \cite{8}
considered one point function of R-R vertex operator on the {\it disk}.
Since the total right + left ghost charge number must add up to $-2$, one is
forced to change the vertex operator in the conventional $(-\frac{1}{2},-%
\frac{1}{2})$ picture to either $(-\frac{1}{2},-\frac{3}{2})$ or $(-\frac{3}{%
2},-\frac{1}{2})$ picture. This inverse picture changing involves, among
other unrelated things, a factor of $k\cdot \Gamma $ , which shifts the
field strength to the potential and gives a strong hint that D-brane carries
the R-R charge. On the other hand, our space-time T-dual R-R zero-norm
states do contain this important $k\cdot \Gamma $ factor as can be seen from
eqs (16) and (17). This again gives a strong support of our space-time
T-dual R-R zero-norm state approach. In the next section, we will see an
even more interesting application of these states.

\section{Dilaton-Axion Symmetries and SL(2,Z) S-duality}

According to section II, for II A theory, we have $A_{(1)},$ $A_{(3)}$
potentials with $d\omega _{(0)}^{(T)}$, $d\omega _{(2)}^{(T)}$ T-dual
zero-norm states and, for II B theory, we have $A_{(2)}$, $A_{(4)}$
potential with $d\omega _{(1)}^{(T)}$, $d\omega _{(3)}^{(T)}$ T-dual
zero-norm states for their symmetry charge parameters. For completeness we
have, in addition, an axion $A_{(0)}\equiv \chi $ in the II B theory. The
corresponding T-dual zero-norm state is naturally identified to be the
constant 0-form $F_{(0)}^{(T)}$, which is Poincare dual to the {\it constant}
10-form $F_{(10)}^{(T)}\equiv d\omega _{(9)}^{(T)}$. So we have the
''symmetry''

\begin{equation}
\chi \rightarrow \chi +F_{(0)}^{(T)}.
\end{equation}
Note that, in eq. (8), there is a constant 10-form field strength $%
G_{(10)}=dA_{(9)}$ which is Poincare dual to the constant 0-form field
strength in II A theory as well. This non-propagating degree of freedom can
be included in the massive type II A supergravity and was conjectured to be
related to the cosmological constant. See the interesting discussion of this
9-form potential $A_{(9)}$ by Polchinski in Ref.\cite{7}. Equation (21) is
consistent with the fact that the axion $\chi $ is defined up to a constant.
The interesting new result here is that we naturally identify this constant
to be $F_{(0)}^{(T)}$.

We now turn to the discussion of NS-NS dilaton $\phi $. Remember we have a
Remaining NS-NS singlet zero-norm state in equation (4). The physical
meaning of this state will be discussed in the following. In reference \cite
{11} each space-time symmetry of the bosonic background field in the first
order WFA can be constructed through a superconformal deformation

\begin{equation}
(T^{(1)}=\overline{T}^{(1)},T_{F}^{(1)},\overline{T}_{F}^{(1)})
\end{equation}
corresponding to a spacetime zero-norm state.

In equation (22), $T_{{}}^{(1)}$($T_{F}^{(1)}$) is the upper component
(lower component) of deformation of the superstress tensor in the first
order WFA when the background field is turned on. $\overline{T}^{(1)},(%
\overline{T}_{F}^{(1)})$ is its anti-holomorphic counterpart. It was shown
that superconformal deformations constructed from zero-norm states in eq.(3)
give the symmetries of graviton and antisymmetric tensor. The superconformal
deformation constructed from the zero-norm state in eq.(4), which was
neglected in the previous discussion., is calculated to be

\begin{eqnarray}
T^{(1)} &=&\overline{T}^{(1)}=\partial _{\mu }\partial _{\nu }\theta
(\partial x^{\mu }+\overleftarrow{\partial _{\lambda }}\psi ^{\lambda }\psi
^{\mu })(\overline{\partial }x^{\nu }+\overleftarrow{\partial _{r}}\overline{%
\psi }^{r}\overline{\psi }^{\nu })  \nonumber \\
&=&\partial _{\mu }\partial _{\nu }\theta \partial x^{\mu }\overline{%
\partial }x^{\nu },  \eqnum{23a}
\end{eqnarray}

\begin{equation}
T_{F}^{(1)}=\frac{1}{2}\partial _{\mu }\partial _{\nu }\theta \overline{%
\partial }x^{\nu }\psi ^{\mu },  \eqnum{23b}
\end{equation}

\begin{equation}
\overline{T}_{F}^{(1)}=\frac{1}{2}\partial _{\mu }\partial _{\nu }\theta
\partial x^{\nu }\overline{\psi }^{\mu }  \eqnum{23c}
\end{equation}
with condition $\Box \theta =constant$, $\Box \equiv \partial ^{\mu
}\partial _{\mu }.$ $\theta (x)$ in eq(23) is the background field
corresponding to the singlet zero-norm state of eq(4). The induced
''symmetry'' is calculated to be

\begin{equation}
\phi \rightarrow \phi +\Box \theta  \eqnum{24}
\end{equation}
and

\begin{equation}
h_{\mu \nu }\rightarrow h_{\mu \nu }+\partial _{\mu }\partial _{\nu }\theta .
\eqnum{25}
\end{equation}
Equation (25) is merely a change of gauge in the linearized graviton and can
be absorbed to the symmetry of the linearized graviton. Equation (24) is the
''symmetry'' of the dilaton. The result that $\Box \theta $ is a constant is
consistent with the fact that $\phi $ appears in the effecive equation of
motion, constructed from vanishing $\sigma -$model $\beta -$function\cite{13}%
$,$ in an overall factor $e^{-2\phi }$ other than differentiated. The
interesting result here is that we identify the constant $\square \theta $
to be the zero-norm state in equation (4). This completes the physical
effects of all massless zero-norm states in the type II string spectrum. The
''symmetries'' presented in equations (21) and (24) were derived in the
first order WFA. They can be broken in the higher order correction. However,
if one tries to generalize the superconformal deformation to {\it second}
order in the WFA, one immediately meets the difficulty of nonperturbative
nonnormalizibility of the 2d $\sigma -%
\mathop{\rm mod}%
$el$,$ and is forced to introduce counterterms which consist of an infinite
number of massive tensor fields.\cite{14} This higher order effect is
related to the stringy physics ($\alpha ^{\prime }\rightarrow \infty $) of
string theory instead of point particle field theory. In fact, it was known
that there exist important stringy bound states called (p,q) string which
consists of p F-strings and q D-strings in the II B theory.\cite{15} The
coupling of axion $\chi $ to (p,q) string, which is an higher order effect
and so can not be seen in our first order WFA, breaks the symmetry in
equation (21) down to integer shifts. On the other hand, it was known that
the symmetry in eq.(24) was broken down to the discrete $\Box \theta \equiv
-2\left\langle \phi \right\rangle $ from the type II B supergravity. If we
define

\begin{equation}
\rho =\chi +ie^{-\phi },\left\langle \rho \right\rangle \equiv \frac{\theta 
}{2\pi }+\frac{i}{g_{s}}\equiv \tau ,  \eqnum{26}
\end{equation}
these two discrete symmetries combine to form the well-known SL(2,Z)
S-duality symmetry of II B string. Note that the nonlinearity of SL(2,Z)
does not appear in our linear WFA. This is a generic feature of WFA in
contract to the usual $\sigma $-model loop($\alpha ^{\prime }$) expansion.
The former contains stringy phenomena (e. g. high energy symmetries) which
can not be derived in the loop expansion scheme, while the latter is
convenient to obtain the low energy effective field theory of the
superstring. This will become clear when one considers the massive states of
the string, which are crucial to make string theory different from the usual
quantum field theory. An immediate application of this Type II B S-duality
is the N=4, d=4 SUSY Yang-Mills S-duality\cite{16}, where dyon with the
electric charge p and magnetic charge q can be interpreted as the end points
of the (p,q) string on the D3-brane. The $\tau $ parameter in this SUSY
gauge theory is defined to be

\begin{equation}
\tau =\frac{\theta _{YM}}{2\pi }+\frac{i}{g_{YM}^{2}},  \eqnum{27}
\end{equation}
and is interpreted to be the constant $\rho $ field of II B string in
equation (26) associated to a stack of D3-branes.

\section{Conclusion}

T-dual R-R zero-norm states motivate the introduction of D-branes into Type
II string theory{\it . }They serve as symmetry charge parameters of R-R
tensor forms. The study in this paper reveals again that all space-time
symmetries, including the {\it discrete }T-duality and S-duality, are
related to the zero-norm states in the spectrum. The unified description of
S and T dualities makes one to speculate that they are all geometric
symmetries (due to the redefinition of string backgrounds) and to conjecture
the existence of a bigger discrete U-duality symmetry\cite{9},\cite{17} and
its relation to the zero-norm state. In fact the SL(2,Z) S-duality of II B
string led Vafa\cite{18} to propose a 12d F-theory, where $\tau $ is the
geometric complex structure modulus of torus T$^{2}$. One can even
generalize this zero-norm state idea to construct new stringy massive
symmetries of string theory. In particular, the existence of some massive
R-R zero-norm states and other evidences make us speculate that string may
carry some massive R-R charges\cite{6}. Another interesting issue is the
identification of D-brane charges with elements of K-theory groups.\cite{19}
How T-dual R-R zero-norm states relate to K-theory groups is an interesting
question to study.

\section{Acknowledgments}

I would like to thank Pei-Ming Ho and Miao Li for comments. This research is
supported by National Science Council of Taiwan, under grant number NSC
89-2112-M-009-006.

\end{document}